# Coupling cavity model for circular cylindrical waveguide with uniform cross section

M.I. Ayzatsky[1], V.V.Mytrochenko

National Science Center Kharkov Institute of Physics and Technology (NSC KIPT),
610108, Kharkov, Ukraine

We developed the general approach that gives possibility to calculate the coupling coefficients for arbitrary chain of resonators without using the great number of eigen functions. For understanding this method and having possibility to control the accuracy of obtained results, we applied this procedure for simplest structure that has the analytical solutions - a circular cylindrical waveguide with uniform cross section.

## 1 Introduction

Development of the new Coupling Cavity Model (CCM) on the base of the Mode Matching Method (MMM) gave possibility to look more deeply into the properties of the inhomogeneous disk loaded waveguides (DLWs) [1,2,3,4,5], especially on the basis of methods that are used for tuning such DLWs [6,7]

In the frame of the CCM such coupling equations can be obtained [1-5]

$$Z_k e_{q_0}^{(k)} = \sum_l e_{q_0}^{(l)} \alpha_{q_0}^{(k,l)},$$

where $e_{q_0}^{(k)}$ - amplitudes of the basis $E_{q_0}$ mode, $Z_k = 1 - \omega^2 / \omega_{q_0}^{(k)2}$, $\omega_{q_0}^{(k)}$ - eigen frequency of this mode in the k-cell, $\alpha_{q_0}^{(k,l)}$ - coupling coefficients that depend on the frequency and the geometric sizes of coupling volumes.

There are some drawbacks in the developed approach for calculating the coupling coefficients $\alpha_{q_0}^{(k,l)}$. First one is the possibility conducting full numerical simulation of nonuniform DLW and obtaining all necessary coupling coefficients for the simple cell geometry only. The CCM, that was developed recently [1-5], can be used only for geometries for which there are analytical expressions of the eigen functions. Secondly, for receiving the necessary accuracy of simulation we must take into account the great number of eigen functions that can be difficult from several reasons (accuracy of Bessel function calculation, etc).

Therefore, it is necessary to develop the general approach that gives possibility to obtain the coupling coefficients for arbitrary chain of resonators without using the great number of eigen functions. We proposed the new method that is based on using only one eigen vector [7,8]. The waveguide is divided into a set of arbitrary volumes. The procedure of calculation of the coupling coefficients consists of two main stages. The first one is the numerical solving of the boundary value problem for the basis mode in each volume and finding its field distribution. The second step is more complicated – numerical solution of Maxwell equations with the magnetic current that is determined by the structure of basis mode. This solution is to be satisfied certain boundary conditions. For understanding this method and having possibility to control the accuracy of obtained results, we applied this procedure for simplest structure that has the analytical solutions - a circular cylindrical waveguide with uniform cross section. For this case, we have to deal with 1-D problem.

The detailed description of application this method to 1-D geometry is given in this article.

---

[1] M.I. Aizatskyi, N.I.Aizatsky; aizatsky@kipt.kharkov.ua

# 1 The Coupling Cavity Model on the base of the Mode Matching Method

The first method to be discussed for obtaining the coupling cavity equations for circular cylindrical waveguide with uniform cross section is the Mode Matching Method

We will assume that all field quantities have a time variation given by $\exp(-i\omega t)$. The electrical charge and current are assumed to be absent in the waveguide. The direction of propagation in the guide is along the z axis. The electromagnetic field is governed by the Maxwell equations

$$\nabla \times \vec{E} = i\omega\mu_0 \vec{H},$$
$$\nabla \times \vec{H} = -i\omega\varepsilon_0\varepsilon \vec{E}.$$
(1.1)

Let's divide the waveguide into the set of arbitrary volumes $V_k$ as shown in Fig. 1 and represent electromagnetic field in each volume $V_k$ as[2]

$$\vec{E} = \sum_q e_q^{(k)} \vec{\mathbb{E}}_q^{(k)}(\vec{r})$$
(1.2)

$$\vec{H} = i\sum_q h_q^{(k)} \vec{\mathbb{H}}_q^{(k)}(\vec{r}), \ \vec{r} \in V_k,$$
(1.3)

where $\vec{\mathbb{E}}_q^{(k)}, \vec{\mathbb{H}}_q^{(k)}$ - solenoidal eigen vectors for the volume $V_k$, $\omega_q^{(k)}$ - eigen frequencies for the volume $V_k$. Vectors $\vec{\mathbb{E}}_q^{(k)}, \vec{\mathbb{H}}_q^{(k)}$ satisfy such equations

$$\nabla \times \vec{\mathbb{E}}_q^{(k)} = i\omega_q^{(k)}\mu_0 \vec{\mathbb{H}}_q^{(k)},$$
$$\nabla \times \vec{\mathbb{H}}_q^{(k)} = -i\omega_q^{(k)}\varepsilon_0\varepsilon \vec{\mathbb{E}}_q^{(k)}, \ \vec{r} \in V_k$$
(1.4)

with boundary conditions $\vec{\mathbb{E}}_{q,\tau}^{(k)} = 0$ at the volume $V_k$ border.

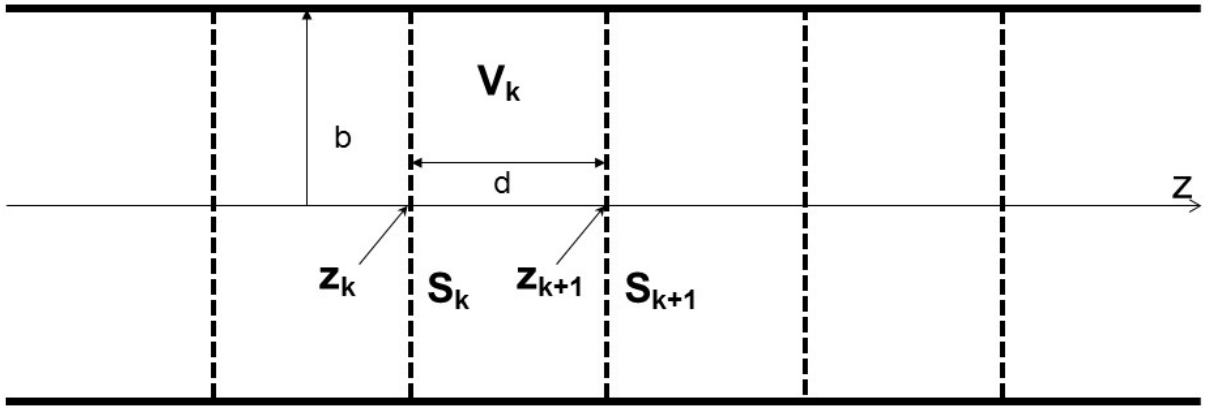

Fig. 1

For TM modes with axial symmetry ($q = (0, m, n)$) the field components may be written as

$$\mathbb{E}_{m,n,z}^{(k)} = J_0\left(\frac{\lambda_m}{b}r\right)\cos\left(\frac{\pi}{d}n z'\right),$$
(1.5)

$$\mathbb{H}_{m,n,\varphi}^{(k)} = -i\omega_{m,n}\frac{\varepsilon_0\varepsilon b}{\lambda_m} J_1\left(\frac{\lambda_m}{b}r\right)\cos\left(\frac{\pi}{d}n z'\right),$$
(1.6)

---

[2] We will consider electromagnetic fields with amplitudes of potential modes equal zero

$$\mathbb{E}_{m,n,r}^{(k)} = \frac{b}{\lambda_m} \frac{\pi n}{d} J_1\left(\frac{\lambda_m}{b} r\right) \sin\left(\frac{\pi}{d} n z'\right) \tag{1.7}$$

$$\omega_{m,n}^2 = \frac{c^2}{\varepsilon}\left\{\left(\frac{\lambda_m}{b}\right)^2 + \left(\frac{\pi n}{d}\right)^2\right\}, \tag{1.8}$$

where $J_0(\lambda_m) = 0$, $z' = z - z_k$, $0 \le z' \le d$.

Eigen oscillation norm is determined as

$$N_{m,n}^{(k)} = \mu_0 \int_{V_k} \vec{\mathbb{H}}_q^{(k)} \vec{\mathbb{H}}_q^{(k)*} dV = \varepsilon_0 |\varepsilon| \int_{V_k} \vec{\mathbb{E}}_q^{(k)} \vec{\mathbb{E}}_q^{(k)*} dV = \frac{\omega_{n,m}^2 b^4}{2c^2 \lambda_m^2} \varepsilon_0 |\varepsilon|^2 \pi d J_1^2(\lambda_m) \sigma_n, \tag{1.9}$$

$$\sigma_n = \begin{cases} 2 & n = 0, \\ 1 & n > 0. \end{cases} \tag{1.10}$$

Amplitudes $e_q^{(k)}$ and $h_q^{(k)}$ can be expressed through the tangential components of the electric fields on the surfaces $S_k$ and $S_{k+1}$

$$\left(\omega_q^{(k)2} - \omega^2\right) e_q^{(k)} = \frac{i \omega_q^{(k)*}}{N_q^{(k)}} \frac{|\varepsilon|}{\varepsilon} \left( \oint_{S_k} [\vec{E}_\tau^{(k)} \vec{\mathbb{H}}_q^{(k)*}] d\vec{S} + \oint_{S_{k+1}} [\vec{E}_\tau^{(k+1)} \vec{\mathbb{H}}_q^{(k)*}] d\vec{S} \right), \tag{1.11}$$

$$h_q^{(k)} = -i \frac{\omega}{\omega_q^{(k)*}} \frac{\varepsilon}{|\varepsilon|} e_q^{(k)} \tag{1.12}$$

Index $\tau$ here and bellow designate tangential components of vector

$$\vec{E}_\tau = \vec{e}_r E_r \tag{1.13}$$

There is a complete orthogonal set of the basis functions $\left\{J_1\left(\frac{\lambda_s}{b} r\right)\right\}$ in the transverse aperture $S_k$ and we can decompose $E_r^{(k)}(r)$ as

$$E_r^{(k)} = \sum_s C_s^{(k)} J_1\left(\frac{\lambda_s}{b} r\right) \tag{1.14}$$

Combining (1.11) and (1.14), we arrive at the expression

$$e_{m,n}^{(k)} = -\frac{2c^2}{\varepsilon \sigma_n d} \frac{\left(-C_m^{(k)} + (-1)^n C_m^{(k+1)}\right)}{\left(\omega_{m,n}^2 - \omega^2\right)} \tag{1.15}$$

From this expression it follows that we may consider the field components with only one radial index $m = m_0$.

Matching of the fields $\vec{H}_\tau$ [3] at the interface $z = z_k$ leads to the following relations

$$H_\varphi(r, z_{k+1} - 0) = i \sum_q h_q^{(k)} \vec{\mathbb{H}}_{q,\varphi}^{(k)}(r, d) = \vec{H}_\tau(\vec{r}_\perp, z_{k+1} + 0) = i \sum_q h_q^{(k+1)} \vec{\mathbb{H}}_{q,\tau}^{(k+1)}(\vec{r}_\perp, 0) \tag{1.16}$$

Decomposing of the left and right hand parts of the equality (1.16) in terms of a complete orthogonal set of the functions $\left\{J_1\left(\frac{\lambda_{s'}}{b} r\right)\right\}$ and using (1.12), we arrive at the result

$$\sum_n e_{m_0,n}^{(k-1)} (-1)^n = \sum_n e_{m_0,n}^{(k)}, \tag{1.17}$$

The traditional next step in the studding of waveguide properties is substituting (1.15) into (1.17). When this is accomplished, we obtain the coupling equations for $C_{m_0}^{(k)}$

$$C_{m_0}^{(k-1)} \Lambda_{m_0}^{(2)} - 2 C_{m_0}^{(k)} \Lambda_{m_0}^{(1)} + C_{m_0}^{(k+1)} \Lambda_{m_0}^{(2)} = 0 \tag{1.18}$$

---

[3] We cannot match electric field as $\vec{\mathbb{E}}_{q,\tau}^{(k)} = 0$ at the volume $V_k$ border

where

$$\Lambda_{m_0}^{(1)} = \frac{2c^2}{d^2} \sum_n \frac{1}{\sigma_n \left(\omega_{m_0,n}^2 - \omega^2\right)} = \frac{cth(d\gamma_{m_0})}{d\gamma_{m_0}}, \qquad (1.19)$$

$$\Lambda_{m_0}^{(2)} = \frac{2c^2}{d^2} \sum_n \frac{(-1)^n}{\sigma_n \left(\omega_{m_0,n}^2 - \omega^2\right)} = \frac{1}{d\gamma_{m_0} sh(d\gamma_{m_0})}, \qquad (1.20)$$

$$\gamma_{m_0} = \sqrt{\frac{\lambda_{m_0}^2}{b^2} - \frac{\omega^2}{c^2}\varepsilon} \qquad (1.21)$$

A difference equation (1.18) has solutions

$$C_{m_0}^{(k)} = C\, p_{1,2}^k \qquad (1.22)$$

where $p_{1,2}$ are the solutions of a characteristic equation

$$p^2 - 2\frac{\Lambda_{m_0}^{(1)}}{\Lambda_{m_0}^{(2)}} p + 1 = 0. \qquad (1.23)$$

Its solutions are

$$p_{1,2} = ch(\gamma_{m_0} d) \pm sh(\gamma_{m_0} d) = \exp(\pm\gamma_{m_0} d). \qquad (1.24)$$

We obtained the expected result: there are two waves (propagating or evanescent ones) in the homogeneous waveguide.

In the CCM we rewrite (1.17) as

$$\sum_{n \neq n_0} e_{m_0,n}^{(k-1)} (-1)^n - \sum_{n \neq n_0} e_{m_0,n}^{(k)} = e_{m_0,n_0}^{(k)} - e_{m_0,n_0}^{(k-1)} (-1)^{n_0}, \qquad (1.25)$$

where $n_0$ is the longitudinal index of a basis mode.

Using (1.15), we obtain the difference equation

$$C_{m_0}^{(k-1)} \Lambda_{m_0,n_0}^{(2)} - 2C_{m_0}^{(k)} \Lambda_{m_0,n_0}^{(1)} + C_{m_0}^{(k+1)} \Lambda_{m_0,n_0}^{(2)} = -(-1)^{n_0} e_{m_0,n_0}^{(k-1)} + e_{m_0,n_0}^{(k)}, \qquad (1.26)$$

with

$$\Lambda_{m_0,n_0}^{(1)} = \frac{2c^2}{d^2} \sum_{n \neq n_0} \frac{1}{\sigma_n \left(\omega_{m_0,n}^2 - \omega^2\right)} = \frac{cth(d\gamma_{m_0})}{d\gamma_{m_0}} - \frac{2}{\sigma_{n_0} \left(\pi^2 n_0^2 + d^2\gamma_{m_0}^2\right)} \qquad (1.27)$$

$$\Lambda_{m_0,n_0}^{(2)} = \frac{2c^2}{d^2} \sum_{n \neq n_0} \frac{(-1)^n}{\sigma_n \left(\omega_{m_0,n}^2 - \omega^2\right)} = \frac{1}{d\gamma_{m_0} sh(d\gamma_{m_0})} - \frac{2(-1)^{n_0}}{\sigma_{n_0} \left(\pi^2 n_0^2 + d^2\gamma_{m_0}^2\right)}, \qquad (1.28)$$

We may divide $C_{m_0}^{(k)}$ into two parts

$$C_{m_0}^{(k)} = C_{m_0}^{(k,1)} + C_{m_0}^{(k,2)}, \qquad (1.29)$$

where $C_{m_0}^{(k,1)}$ is the solution of the inhomogeneous difference equation

$$C_{m_0}^{(k-1,1)} \Lambda_{m_0,n_0}^{(2)} - 2C_{m_0}^{(k,1)} \Lambda_{m_0,n_0}^{(1)} + C_{m_0}^{(k+1,1)} \Lambda_{m_0,n_0}^{(2)} = -(-1)^{n_0} e_{m_0,n_0}^{(k-1)} = -(-1)^{n_0} \sum_l e_{m_0,n_0}^{(l)} \delta_{k-1,l}, \qquad (1.30)$$

and $C_{m_0}^{(k,2)}$

$$C_{m_0}^{(k-1,2)} \Lambda_{m_0,n_0}^{(2)} - 2C_{m_0}^{(k,2)} \Lambda_{m_0,n_0}^{(1)} + C_{m_0}^{(k+1,2)} \Lambda_{m_0,n_0}^{(2)} = e_{m_0,n_0}^{(k)} = \sum_l e_{m_0,n_0}^{(l)} \delta_{k,l}, \qquad (1.31)$$

Solution of the equations (1.31) and (1.30) may be written with using a Green function $\tilde{C}_{m_0}^{(k,l)}$. Then the solution of the equation (1.26) takes the form

$$C_{m_0}^{(k)} = \sum_{l=-\infty}^{l=\infty} e_{m_0,n_0}^{(l)} C_{m_0}^{(k,l)}, \qquad (1.32)$$

where

$$C_{m_0}^{(k,l)} = \left\{\tilde{C}_{m_0}^{(k,l)} - (-1)^{n_0} \tilde{C}_{m_0}^{(k,l+1)}\right\}, \qquad (1.33)$$

The Green function $\tilde{C}_{m_0}^{(k,l)}$ is the solution of the difference equation

$$\tilde{C}_{m_0}^{(k-1,l)}\Lambda_{m_0,n_0}^{(2)} - 2\tilde{C}_{m_0}^{(k,l)}\Lambda_{m_0,n_0}^{(1)} + \tilde{C}_{m_0}^{(k+1,l)}\Lambda_{m_0,n_0}^{(2)} = \delta_{k,l}, \qquad (1.34)$$

where $\delta_{k,l}$ is the Kronecker symbol.

Solution of the equation (1.34) is

$$\tilde{C}_{m_0}^{(k,l)} = \frac{1}{\Lambda^{(2)}(p_2 - p_1)}\begin{cases} p_1^{k-l}, & k \leq l \\ p_2^{k-l}, & k \geq l+1 \end{cases}, \qquad (1.35)$$

where

$$p_{1,2} = \frac{\Lambda_{m_0,n_0}^{(1)}}{\Lambda_{m_0,n_0}^{(2)}} \pm \sqrt{\left(\frac{\Lambda_{m_0,n_0}^{(1)}}{\Lambda_{m_0,n_0}^{(2)}}\right)^2 - 1}. \qquad (1.36)$$

It is necessary to specify the signs in (1.36). The specification is obtained from the condition that the solution (1.35) vanishes at the infinity. The signs is to be chosen in such way that $|p_2| < 1$ ($p_1 p_2 = 1$).

It follows from (1.34) that in the case of homogeneous waveguide the Green function $\tilde{C}_{m_0}^{(k,l)}$ satisfy the condition

$$\tilde{C}^{(s+S,l+S)} = \tilde{C}^{(s,l)} \qquad (1.37)$$

Substituting (1.35) into (1.33), we have

$$C_{m_0}^{(k,l)} = \frac{1}{\Lambda_{m_0}^{(2)}(p_2 - p_1)}\begin{cases} \{1-(-1)^{n_0} p_1\} p_2^{k-l}, & k \geq l+1 \\ \{1-(-1)^{n_0} p_2\} p_1^{k-l}, & k \leq l \end{cases} \qquad (1.38)$$

Finally, we can write the coefficients $C_{m_0}^{(k)}$ as

$$C_{m_0}^{(k)} = \frac{1}{\{p_2 - p_1\}\Lambda_{m_0,n_0}^{(2)}}\left[\{1-(-1)^{n_0} p_1\}\sum_{l=-\infty}^{l=k-1} p_2^{k-l} e_{m_0,n_0}^{(l)} + \{1-(-1)^{n_0} p_2\}\sum_{l=k}^{l=\infty} p_1^{k-l} e_{m_0,n_0}^{(l)}\right], \qquad (1.39)$$

Equations for the amplitudes of the basis mode take then the form

$$\left(\omega_{m_0,n_0}^2 - \omega^2\right)e_{m_0,n_0}^{(k)} = \frac{2c^2}{\varepsilon\sigma_{n_0} d^2}\left(C_{m_0}^{(k)} - (-1)^n C_{m_0}^{(k+1)}\right) =$$
$$= \frac{2c^2}{\varepsilon\sigma_{n_0} d^2}\sum_{l=-\infty}^{l=\infty} e_{m_0,n_0}^{(l)}\left(C_{m_0}^{(k,l)} - (-1)^{n_0} C_{m_0}^{(k+1,l)}\right) = \omega_{m_0,n_0}^2 \sum_{l} e_{m_0,n_0}^{(l)} \alpha_{m_0,n_0}^{(k,l)} \qquad (1.40)$$

where the coupling coefficients $\alpha_{q_0}^{(k,l)}$

$$\alpha_{m_0,n_0}^{k,l} = \frac{2c^2}{\varepsilon\sigma_{n_0} d\, \omega_{m_0,n_0}^2}\left(C_{m_0}^{(k,l)} - (-1)^{n_0} C_{m_0}^{(k+1,l)}\right) =$$
$$= \frac{2c^2}{\varepsilon\sigma_{n_0} d\, \omega_{m_0,n_0}^2}\left(\tilde{C}_{m_0}^{(k,l)} - (-1)^{n_0}\tilde{C}_{m_0}^{(k,l+1)} - (-1)^{n_0}\tilde{C}_{m_0}^{(k+1,l)} + \tilde{C}_{m_0}^{(k+1,l+1)}\right) = \qquad (1.41)$$
$$= \frac{2c^2}{\varepsilon\sigma_{n_0} d^2\omega_{m_0,n_0}^2(p_2 - p_1)\Lambda_{m_0,n_0}^{(2)}}\begin{cases} 2\{1-(-1)^{n_0} p_2\}, & l = k \\ \{2-(-1)^{n_0} p_1 -(-1)^{n_0} p_2\} p_2^{|k-l|}, & l \neq k \end{cases}$$

Let us note that

$$\alpha_{m_0,n_0}^{k,l} = \alpha_{m_0,n_0}^{|k-l|} \qquad (1.42)$$

It is convenient to rewrite (1.40) as

$$\left(\omega_{m_0,n_0}^2 - \omega^2\right)e_{m_0,n_0}^{(k)} = \omega_{m_0,n_0}^2 \sum_l e_{m_0,n_0}^{(l)} \alpha_{m_0,n_0}^{(k,l)} =$$
$$= \omega_{m_0,n_0}^2 \sum_{l'} e_{m_0,n_0}^{(k+l)} \alpha_{m_0,n_0}^{(k,k+l)} \quad , \tag{1.43}$$

The coupling coefficients $\alpha_{q_0}^{(k,l)}$ depend on geometry of the waveguide under consideration and the frequency $\omega$.

Electromagnetic fields may be found by substituting expression for amplitudes

$$e_{m_0,n}^{(k)} = \frac{2c^2}{\varepsilon\sigma_n d} \sum_{l=-\infty}^{l=\infty} e_{m_0,n_0}^{(l)} \frac{\left\{\tilde{C}_{m_0}^{(k,l)} - (-1)^{n_0}\tilde{C}_{m_0}^{(k,l+1)} - (-1)^{n_0}\tilde{C}_{m_0}^{(k+1,l)} + \tilde{C}_{m_0}^{(k+1,l+1)}\right\}}{\left(\omega_{m_0,n}^2 - \omega^2\right)} \tag{1.44}$$

into the expansions (1.2) and (1.3). For example,

$$E_z = \sum_q e_q^{(k)}\,\mathbb{E}_{q,z}^{(k)}(\vec{r}) = e_{q_0}^{(k)}\,\mathbb{E}_{q_0,z}^{(k)}(\vec{r}) + \sum_{q \neq q_0} e_q^{(k)}\,\mathbb{E}_{q,z}^{(k)}(\vec{r}) =$$
$$= -C\exp(k\gamma_{m_0}d)\exp(\gamma_{m_0}z')J_0\left(\frac{\lambda_{m_0}}{a}r\right)\frac{1}{\varepsilon d \gamma_{m_0}\Lambda^{(2)}}\frac{p_2\left\{\exp(-\gamma_{m_0}d)(-1)^{n_0}-1\right\}}{1-2p_2\operatorname{ch}(\gamma_{m_0}d)+p_2^2} = \tag{1.45}$$
$$= \tilde{C}\exp(k\gamma_{m_0}d)\exp(\gamma_{m_0}z')J_0\left(\frac{\lambda_{m_0}}{a}r\right) = \tilde{C}\exp(\gamma_{m_0}z)J_0\left(\frac{\lambda_{m_0}}{a}r\right)$$

This result coincides with the known field pattern of the longitudinal electric field in the homogeneous circular waveguide.

Coupling equations (1.40) have the solution of the form $e_{m_0,n_0}^{(k)} \sim \exp(ik\varphi)$, where $\varphi$ is determined by the dispersive equation

$$D(\omega,\varphi) = \frac{4c^2}{\sigma_{n_0}d^2\varepsilon(p_2-p_1)\left(\omega_{m_0,n_0}^2-\omega^2\right)\Lambda_{m_0,n_0}^{(2)}}\left(\begin{array}{c}\left\{1-(-1)^{n_0}p_2\right\}+\\ +\left[\left\{2-(-1)^{n_0}p_1-(-1)^{n_0}p_2\right\}\right]\sum_{l=1}^{l=\infty}p_2^l\cos(l\varphi)\end{array}\right)-1=$$
$$= \frac{4c^2\left(1-p_2^2\right)}{\sigma_{n_0}d^2\varepsilon\{p_1-p_2\}\left(\omega_{m_0,n_0}^2-\omega^2\right)\Lambda_{m_0,n_0}^{(2)}}\frac{1-(-1)^{n_0}\cos(\varphi)}{1-2\cos(\varphi)p_2+p_2^2}-1=0$$
$$\tag{1.46}$$

It can be shown that this equation has the solution of the form $\varphi = \pm i\gamma_{m_0}d$ for arbitrary values of $m_0$ and $n_0$.

As an example, in Fig. 2 and Fig. 3 we present the dependencies of $\log\{abs(p_1)\}, \log\{abs(p_2)\}$ and $\operatorname{Real}\{D(\Omega,\varphi=\pm i\gamma_{m_0}d)\}$ on dimensionless frequency $\Omega = \frac{\omega b}{c}$ for ($m_0=1$, $n_0=0$.) and ($m_0=3$, $n_0=2$). It is seen that for all frequencies $\varphi = \pm i\gamma_{m_0}d$ is the solution of the dispersive equation (1.46).

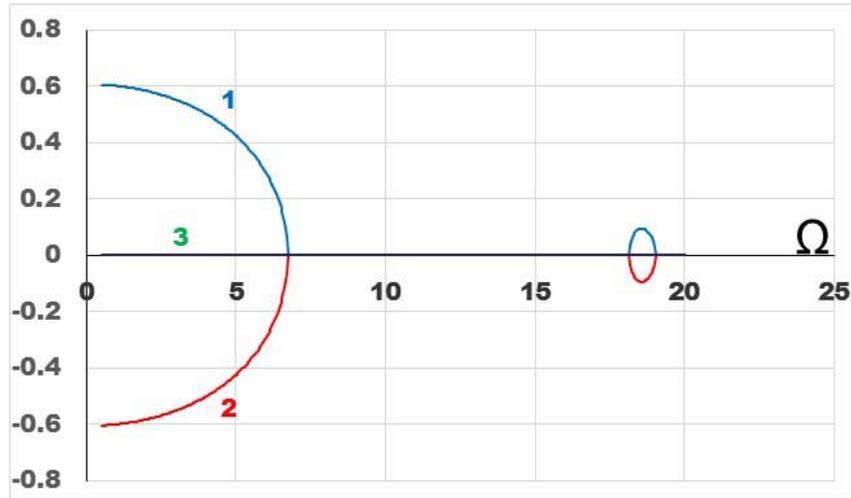

Fig. 2 The dependencies of $\log_{10}\{abs(p_1)\}$ (1), $\log_{10}\{abs(p_2)\}$ (2) and $\mathrm{Abs}\{D(\Omega, \varphi = \pm i\gamma_{m_0} d)\}$ (3) on dimensionless frequency $\Omega = \dfrac{\omega b}{c}$ for $m_0 = 1$ and $n_0 = 0$ ($b/d = 2$, $\varepsilon = 1 - 1E - 6i$ )

We also see that there are the intervals of frequency where the difference $(|p_2| - 1)$ is small[4]. In this case the convergence of the sum in the right-hand part of (1.43) is slow as the coupling coefficients $\alpha_{m_0,n_0}^{(k,k+l)}$ decrease as $p_2^{|l|}$ when $l \to \infty$ (see (1.41)).

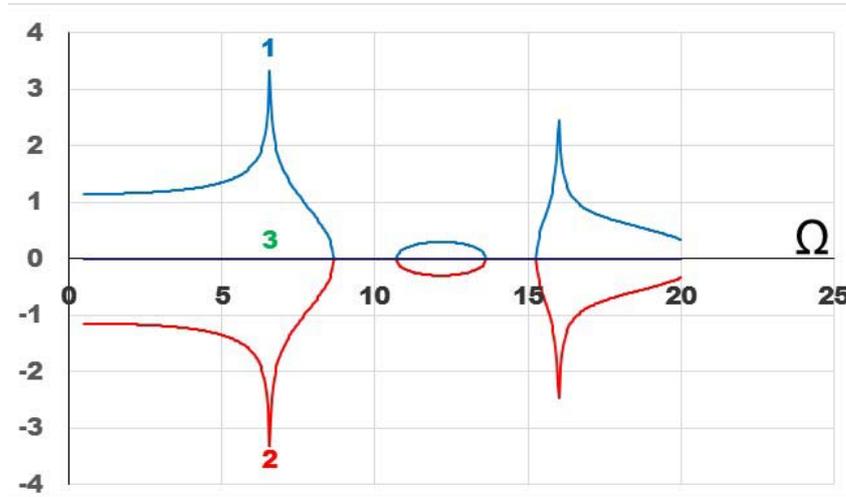

Fig. 3 The dependencies of $\log_{10}\{abs(p_1)\}$ (1), $\log_{10}\{abs(p_2)\}$ (2) and $\mathrm{Abs}\{D(\Omega, \varphi = \pm i\gamma_{m_0} d)\}$ (3) on dimensionless frequency $\Omega = \dfrac{\omega b}{c}$ for $m_0 = 3$ and $n_0 = 2$ ($b/d = 2$, $\varepsilon = 1 - 1E - 6i$ )

## 2 The Coupling Cavity Model on the base of one eigen mode

### 2.1 General description

In this approach we also divide the waveguide into the set of arbitrary volumes $V_k$ as shown in Fig. 1 and represent electromagnetic field in $V_k$ as

$$\vec{E} = \vec{\tilde{E}} + e_{q_0}^{(k)} \vec{\mathbb{E}}_{q_0}^{(k)}, \quad \vec{r} \in V_k \tag{1.47}$$

$$\vec{H} = \vec{\tilde{H}} + \frac{\omega}{\omega_{q_0}^{(k)*}} \frac{\varepsilon}{|\varepsilon|} e_{q_0}^{(k)} \vec{\mathbb{H}}_{q_0}^{(k)}, \quad \vec{r} \in V_k \tag{1.48}$$

where $q_0$ is the index of the mode that we will consider as the basis one. We will also suppose that amplitudes of the basis mode $e_{q_0}^{(k)}$ can be expressed through the tangential components of the electric fields on the surfaces $S_k$ and $S_{k+1}$

$$\left( \omega_q^{(k)2} - \omega^2 \right) e_{q_0}^{(k)} = \frac{i\omega_{q_0}^{(k)*}}{N_{q_0}^{(k)}} \frac{|\varepsilon|}{\varepsilon} \left( \oint_{S_k} [\vec{E}_\tau^{(k)} \vec{\mathbb{H}}_{q_0}^{(k)*}] d\vec{S} + \oint_{S_{k+1}} [\vec{E}_\tau^{(k+1)} \vec{\mathbb{H}}_{q_0}^{(k)*}] d\vec{S} \right), \tag{1.49}$$

---

[4] We introduce a small imaginary part into $\varepsilon$ that always gives $|p_2| < 1$

Substituting (1.47) and (1.48) into the Maxwell equations (1.1), we obtain

$$\nabla \times \vec{\tilde{E}} - i\omega\mu_0 \vec{\tilde{H}} = i\mu_0 \frac{\omega^2 - \omega_{q_0}^{(k)2}}{\omega_{q_0}^{(k)*}} \frac{\varepsilon}{|\varepsilon|} e_{q_0}^{(k)} \vec{\mathbb{H}}_{q_0}^{(k)} = -\vec{j}_{mag}^{(k)}, \quad \vec{r} \in V_k,$$

$$\nabla \times \vec{\tilde{H}} + i\omega\varepsilon_0 \vec{\tilde{E}} = 0. \tag{1.50}$$

Making use of (1.49), we have

$$\nabla \times \vec{\tilde{E}} - i\omega\mu_0 \vec{\tilde{H}} = \frac{\mu_0}{N_{q_0}^{(k)}} \left( \oint_{S_k} [\vec{\tilde{E}}_\tau^{(k)} \vec{\mathbb{H}}_{q_0}^{(k)*}] d\vec{S} + \oint_{S_{k+1}} [\vec{\tilde{E}}_\tau^{(k+1)} \vec{\mathbb{H}}_{q_0}^{(k)*}] d\vec{S} \right) \vec{\mathbb{H}}_{q_0}^{(k)} = -\vec{j}_{mag}^{(k)},$$

$$\nabla \times \vec{\tilde{H}} + i\omega\varepsilon_0 \vec{\tilde{E}} = 0 \quad \vec{r} \in V_k. \tag{1.51}$$

From (1.51) it f0llows that the fields $\vec{\tilde{E}}, \vec{\tilde{H}}$ satisfy such conditions

$$\int_{V_k} \vec{\tilde{E}}^{(l)} \vec{\mathbb{E}}_{q_0}^{(k)} dV = 0 \tag{1.52}$$

$$\int_{V_k} \vec{\tilde{H}}^{(l)} \vec{\mathbb{H}}_{q_0}^{(k)} dV = 0 \tag{1.53}$$

Indeed, if we decompose the fields $\vec{\tilde{E}}, \vec{\tilde{H}}$ in terms of eigen modes

$$\vec{\tilde{E}} = \sum_q \tilde{e}_q^{(k)} \vec{\mathbb{E}}_q^{(k)}, \quad z_k < z < z_{k+1}, \tag{1.54}$$

$$\vec{\tilde{H}} = i\sum_q \tilde{h}_q^{(k)} \vec{\mathbb{H}}_q^{(k)}(\vec{r}), \quad z_k < z < z_{k+1}, \tag{1.55}$$

then we obtain

$$\frac{\varepsilon}{|\varepsilon|} \frac{\left( \omega_q^{(s)2} - \omega^2 \right) N_q^{(s)}}{i\omega_q^{(s)*}} \tilde{e}_q^{(s)} = \left( \oint_{S_k} [\vec{\tilde{E}}_\tau^{(k)} \vec{\mathbb{H}}_q^{(k)*}] d\vec{S} + \oint_{S_{k+1}} [\vec{\tilde{E}}_\tau^{(k+1)} \vec{\mathbb{H}}_q^{(k)*}] d\vec{S} \right) + \int_V \vec{j}_{mag}^{(k)} \vec{\mathbb{H}}_q^{(s)*} dV, \tag{1.56}$$

The right hand part of (1.56) we can rewrite as

$$\left( \oint_{S_k} [\vec{\tilde{E}}_\tau^{(k)} \vec{\mathbb{H}}_q^{(k)*}] d\vec{S} + \oint_{S_{k+1}} [\vec{\tilde{E}}_\tau^{(k+1)} \vec{\mathbb{H}}_q^{(k)*}] d\vec{S} \right) + \int_V \vec{j}_{mag}^{(k)} \vec{\mathbb{H}}_q^{(s)*} dV =$$

$$= \begin{cases} \left( \oint_{S_k} [\vec{\tilde{E}}_\tau^{(k)} \vec{\mathbb{H}}_q^{(k)*}] d\vec{S} + \oint_{S_{k+1}} [\vec{\tilde{E}}_\tau^{(k+1)} \vec{\mathbb{H}}_q^{(k)*}] d\vec{S} \right), & q \neq q_0 \\ 0, & q = q_0 \end{cases} \tag{1.57}$$

Therefore, there is no terms with $q = q_0$ in the sums (1.54),(1.55), from which it follows that conditions (1.52) and (1.53) are fulfilled.

99

$\vec{\mathbb{E}}^{(k)}_{q_0,\tau} = 0$ at the volume borders, then $\vec{\tilde{E}}_\tau$ is a continues vector field throughout the waveguide. $\tilde{E}_z$ and $\vec{\tilde{H}}$ are not a continues ones at the borders between $V_{k-1}$, $V_k$.

At the surfaces $S_k$ and $S_{k+1}$ (that are the borders between $V_{k-1}$, $V_k$ and $V_k$, $V_{k+1}$) the continuity conditions on $\vec{H}$ may be stated as

$$\vec{\tilde{H}}(z = z_k - 0) + \frac{\omega}{\omega^{(k-1)*}_{q_0}} \frac{\varepsilon}{|\varepsilon|} e^{(k-1)}_{q_0} \vec{\mathbb{H}}^{(k-1)}_{q_0}(z = z_k) = \vec{\tilde{H}}(z = z_k + 0) + \frac{\omega}{\omega^{(k)*}_{q_0}} \frac{\varepsilon}{|\varepsilon|} e^{(k)}_{q_0} \vec{\mathbb{H}}^{(k)}_{q_0}(z = z_k),$$

$$\vec{\tilde{H}}(z = z_{k+1} - 0) + \frac{\omega}{\omega^{(k)*}_{q_0}} \frac{\varepsilon}{|\varepsilon|} e^{(k)}_{q_0} \vec{\mathbb{H}}^{(k)}_{q_0}(z = z_{k+1}) = \vec{\tilde{H}}(z = z_{k+1} + 0) + \frac{\omega}{\omega^{(k+1)*}_{q_0}} \frac{\varepsilon}{|\varepsilon|} e^{(k+1)}_{q_0} \vec{\mathbb{H}}^{(k+1)}_{q_0}(z = z_{k+1}),$$

(1.58)

These conditions are linear with respect to the amplitudes of the basis mode $e^{(k)}_{q_0}$, so we can represent the fields $\vec{\tilde{E}}$, $\vec{\tilde{H}}$ as

$$\vec{\tilde{E}} = \sum_{l=-\infty}^{l=\infty} e^{(l)}_{q_0} \vec{\tilde{E}}^{(l)},$$

$$\vec{\tilde{H}} = \sum_{l=-\infty}^{l=\infty} e^{(l)}_{q_0} \vec{\tilde{H}}^{(l)}$$

(1.59)

The fields $\vec{\tilde{E}}^{(l)}$, $\vec{\tilde{H}}^{(l)}$ are the solution of such equations

$$\nabla \times \vec{\tilde{E}}^{(l)} - i\omega\mu_0 \vec{\tilde{H}}^{(l)} = \frac{\mu_0}{N^{(k)}_{q_0}} \left( \oint_{S_k} [\vec{\tilde{E}}^{(k,l)}_\tau \vec{\mathbb{H}}^{(k)*}_{q_0}] d\vec{S} + \oint_{S_{k+1}} [\vec{\tilde{E}}^{(k+1,l)}_\tau \vec{\mathbb{H}}^{(k)*}_{q_0}] d\vec{S} \right) \vec{\mathbb{H}}^{(k)}_{q_0},$$ (1.60)

$$\nabla \times \vec{\tilde{H}}^{(l)} + i\omega\varepsilon_0 \vec{\tilde{E}}^{(l)} = 0, \quad \vec{r} \in V_k.$$

with the "continuity" conditions at the surfaces $S_k$ and $S_{k+1}$

$$\vec{\tilde{H}}^{(l)}(z = z_k - 0) = \vec{\tilde{H}}^{(l)}(z = z_k + 0) + \frac{\omega}{\omega^{(k)*}_{q_0}} \frac{\varepsilon}{|\varepsilon|} \delta_{k,l} \vec{\mathbb{H}}^{(k)}_{q_0}(z = z_k),$$

$$\vec{\tilde{H}}^{(l)}(z = z_{k+1} - 0) + \frac{\omega}{\omega^{(k)*}_{q_0}} \frac{\varepsilon}{|\varepsilon|} \delta_{k,l} \vec{\mathbb{H}}^{(k)}_{q_0}(z = z_{k+1}) = \vec{\tilde{H}}^{(l)}(z = z_{k+1} + 0),$$

(1.61)

where $\delta_{k,l}$ is the Kronecker symbol.

Equations for the amplitudes of the basis mode (1.49) take then the form

$$\left(\omega^{(k)2}_{q_0} - \omega^2\right) e^{(k)}_{q_0} = \omega^{(k)2}_{q_0} \sum_l e^{(k+l)}_{q_0} \alpha^{(k,k+l)}_{q_0},$$ (1.62)

where the coupling coefficients $\alpha^{(k,l)}_{q_0}$

$$\alpha^{(k,k+l)}_{q_0} = \frac{i\omega^{(k)*}_{q_0}}{\omega^{(k)2}_{q_0} N^{(k)}_{q_0}} \frac{|\varepsilon|}{\varepsilon} \left( \oint_{S_k} [\vec{\tilde{E}}^{(k,k+l)}_\tau \vec{\mathbb{H}}^{(k)*}_{q_0}] d\vec{S} + \oint_{S_{k+1}} [\vec{\tilde{E}}^{(k+1,k+l)}_\tau \vec{\mathbb{H}}^{(k)*}_{q_0}] d\vec{S} \right)$$ (1.63)

If we find vector functions $\vec{\tilde{E}}^{(k,l)}_\tau$ that are the solution of equations (1.60) and fulfilled the continuity conditions (1.61), we can calculate the coupling coefficients $\alpha^{(k,l)}_{q_0}$.

### 2.2 Circular cylindrical waveguide with uniform cross section

In the case of circular cylindrical waveguide with uniform cross section we may represent field components as

$$\tilde{E}_z^{(l)} = \mathcal{E}_z^{(l)}(z) J_0\left(\frac{\lambda_{m_0}}{a} r\right), \tag{1.64}$$

$$\tilde{H}_\varphi^{(l)} = -i\omega \frac{b}{\lambda_{m_0}} \varepsilon_0 \mathcal{H}_\varphi^{(l)}(z) J_1\left(\frac{\lambda_{m_0}}{b} r\right), \tag{1.65}$$

$$\tilde{E}_r^{(l)} = \frac{b}{d\lambda_{m_0}} \mathcal{E}_r^{(l)}(z) J_1\left(\frac{\lambda_{m_0}}{b} r\right) \tag{1.66}$$

$$\tilde{E}_r^{(l)}(z = z_k) = C^{(k,l)} \frac{b}{d\lambda_{m_0}} J_1\left(\frac{\lambda_{m_0}}{b} r\right) \tag{1.67}$$

Then equations (1.60) transform into

$$\frac{\lambda_s}{b} \mathcal{E}_z^{(l)} + \frac{b}{d\lambda_s} \frac{\partial \mathcal{E}_r^{(l)}}{\partial z} =$$
$$= \omega^2 \frac{b}{c^2 \lambda_s} \mathcal{H}_\varphi^{(l)} - \frac{2}{\sigma_n d^2} \frac{b}{\lambda_s} \left(-C^{(k,l)} + (-1)^{n_0} C^{(k+1,l)}\right) \cos\left\{\frac{\pi}{d} n_0 (z - z_k)\right\}, \tag{1.68}$$

$$\mathcal{E}_r^{(l)} = -d \frac{\partial \mathcal{H}_\varphi^{(l)}}{\partial z}, \tag{1.69}$$

$$\mathcal{H}_\varphi^{(l)} = \mathcal{E}_z^{(l)}. \tag{1.70}$$

Combining (1.68),(1.69) and (1.70), we arrive at the equation for magnetic field $\mathcal{H}_\varphi^{(l)}$

$$\frac{d^2 \mathcal{H}_\varphi^{(l)}}{dz^2} - \gamma_{m_0}^2 \mathcal{H}_\varphi^{(l)} = -\frac{2}{\sigma_{n_0} d^2} \left(-C^{(k,l)} + (-1)^{n_0} C^{(k+1,l)}\right) \cos\left\{\frac{\pi}{d} n_0 (z - z_k)\right\} \tag{1.71}$$

where

$$C^{(k,l)} = -d \left.\frac{d\mathcal{H}_\varphi^{(l)}}{dz}\right|_{z=z_k} \tag{1.72}$$

The "continuity" conditions at $z = z_k$ and $z = z_{k+1}$ for the magnetic field $\mathcal{H}_\varphi^{(l)}$ are

$$\mathcal{H}_\varphi^{(l)}(z = z_k - 0) = \mathcal{H}_\varphi^{(l)}(z = z_k + 0) + \delta_{k,l}$$
$$\mathcal{H}_\varphi^{(l)}(z = z_{k+1} - 0) + (-1)^{n_0} \delta_{k,l} = \mathcal{H}_\varphi^{(l)}(z = z_{k+1} + 0) \tag{1.73}$$

As the problem under consideration is linear with respect to the constants $C^{(k,l)}$ and $C^{(k+1,l)}$, we can write

$$\mathcal{H}_\varphi^{(l)}(kd + z') = u^{(1)}(z') C^{(k,l)} + u^{(2)}(z') C^{(k+1,l)}, \; 0 < z' < d. \tag{1.74}$$

The functions $u^{(1)}$ and $u^{(2)}$ are the solution of the two-point boundary value problems

$$\frac{d^2 u^{\binom{1}{2}}}{dz'^2} - \gamma_{m_0}^2 u^{\binom{1}{2}} = -\frac{2}{d^2 \sigma_{n_0}} \cos\left\{\pi n_0 \frac{z'}{d}\right\} \begin{cases} -1 \\ (-1)^{n_0} \end{cases}, \; 0 < z' \le d \tag{1.75}$$

$$\left.\frac{\partial u^{\binom{1}{2}}}{\partial z}\right|_{z=0} = \begin{cases} -d \\ 0 \end{cases} \tag{1.76}$$

$$\left.\frac{\partial u^{\binom{1}{2}}}{\partial z}\right|_{z=d} = \begin{cases} 0 \\ -d \end{cases} \tag{1.77}$$

Substituting (1.74) into (1.73), we obtain

$$u^{(1)}(d) C^{(k-1,l)} + \left\{u^{(2)}(d) - u^{(1)}(0)\right\} C^{(k,l)} - u^{(2)}(0) C^{(k+1,l)} = \delta_{k,l} - (-1)^{n_0} \delta_{k-1,l} \tag{1.78}$$

### 2.1.1 Exact solution

The equations (1.75) have the exact solutions

$$u^{(1)} = \frac{ch\{\gamma_{m_0}(z'-d)\}}{\gamma_{m_0} d \, sh(\gamma_{m_0} d)} + \frac{2c^2}{\sigma_{n_0} d^2 \{\omega^2 - \omega^2_{m_0,n_0}\}} \cos\left(\frac{\pi}{d} n_0 z'\right), \quad (1.79)$$

$$u^{(2)} = -\frac{ch(\gamma_{m_0} z')}{\gamma_{m_0} d \, sh(\gamma_{m_0} d)} - \frac{2c^2(-1)^{n_0}}{\sigma_{n_0} d^2 \{\omega^2 - \omega^2_{m_0,n_0}\}} \cos\left(\frac{\pi}{d} n_0 z'\right). \quad (1.80)$$

It is useful to note that

$$u^{(1)}(0) = \Lambda^{(1)}_{m_0,n_0}, \; u^{(1)}(d) = \Lambda^{(2)}_{m_0,n_0} \quad (1.81)$$

$$u^{(1)}(0) = -\Lambda^{(2)}_{m_0,n_0}, \; u^{(1)}(d) = -\Lambda^{(1)}_{m_0,n_0} \quad (1.82)$$

and (1.78) takes the form

$$C^{(k-1,l)}_{m_0} \Lambda^{(2)}_{m_0,n_0} - 2C^{(k,l)}_{m_0} \Lambda^{(1)}_{m_0,n_0} + C^{(k+1,l)}_{m_0} \Lambda^{(2)}_{m_0,n_0} = \delta_{k,l} - (-1)^{n_0} \delta_{k-1,l} \quad (1.83)$$

$C^{(k,l)}_{m_0}$ can be expressed through the Green function $\tilde{C}^{(k,l)}_{m_0}$ that was introduced in Section 1 (see (1.34))

$$C^{(k,l)}_{m_0} = \{\tilde{C}^{(k,l)}_{m_0} - (-1)^{n_0} \tilde{C}^{(k,l+1)}_{m_0}\}, \quad (1.84)$$

This expression coincide with one (1.33) that was obtained on the base of the MMM method.

Using (1.79) and (1.80), we can write

$$\mathcal{H}^{(l)}_\varphi = -\frac{C^{(k+1,l)} ch\{\gamma_{m_0}(z-z_k)\} - C^{(k,l)} ch\{\gamma_{m_0}(z-z_k-d)\}}{\gamma_{m_0} d \, sh(\gamma_{m_0} d)} - \\ -\frac{2c^2}{\sigma_{n_0} d^2 \{\omega^2 - \omega^2_{m_0,n_0}\}} \left(-C^{(k,l)} + (-1)^{n_0} C^{(k+1,l)}\right) \cos\left\{\frac{\pi}{d} n_0 (z-z_k)\right\} \quad (1.85)$$

The electric field component is given by

$$\mathcal{E}^{(l)}_r = -d\frac{\partial \mathcal{H}^{(l)}_\varphi}{\partial z} = \frac{C^{(k+1,l)} sh\{\gamma_s(z-z_k)\} - C^{(k,l)} sh\{\gamma_s(z-z_k-d)\}}{sh(\gamma_{m_0} d)} - \\ -\frac{2c^2}{\sigma_{n_0} d \{\omega^2 - \omega^2_{m_0,n_0}\}} \frac{\pi}{d} n_0 \left(-C^{(k,l)} + (-1)^{n_0} C^{(k+1,l)}\right) \sin\left\{\frac{\pi}{d} n_0 (z-z_k)\right\} \quad (1.86)$$

It may be shown that

$$\mathcal{E}^{(l)}_r \bigg|_{z=z_k} = C^{(k,l)} \quad (1.87)$$

and

$$\int_{z_k}^{z_{k+1}} dz \mathcal{H}^{(l)}_\varphi \cos\left\{\frac{\pi}{d} n_0 (z-z_k)\right\} = 0 \quad (1.88)$$

The appropriate set of equations for the amplitudes of the basis modes was given in (1.43) and is repeated here for the sake of convenience

$$\left(\omega^2_{m_0,n_0} - \omega^2\right) e^{(k)}_{m_0,n_0} = \omega^2_{m_0,n_0} \sum_{l'} e^{(k+l')}_{m_0,n_0} \alpha^{(k,k+l')}_{m_0,n_0}, \quad (1.89)$$

Expression for the coupling coefficients $\alpha^{(k,k+l)}_{q_0}$ with using (1.37) transforms into

$$\alpha_{m_0,n_0}^{k,k+l'} = \frac{2c^2}{\varepsilon \sigma_{n_0} d \, \omega_{m_0,n_0}^2} \left( C_{m_0}^{(k,k+l')} - (-1)^{n_0} C_{m_0}^{(k+1,k+l')} \right) =$$
$$= \frac{2c^2}{\varepsilon \sigma_{n_0} d \, \omega_{m_0,n_0}^2} \left( C_{m_0}^{(k-l',k)} - (-1)^{n_0} C_{m_0}^{(k-l'+1,k)} \right) \quad (1.90)$$

where $k$ is a fixed integer (number of the volume under consideration). As the waveguide is uniform, the coupling coefficients $\alpha_{q_0}^{(k,k+l)}$ depend only on $l'$ (difference of volume numbers). It follows from (1.90) that in the equation (1.71) and the conditions (1.73) we may deal only with one value of index $l$.

### 2.1.2 Numerical solution

One of the main aims of this work is to illustrate that proposed approach [8] can be transformed into the numerical procedure that can be realized on the base of developed numerical methods and estimate the possible accuracy.

We may discretize (1.75) using a finite difference method on the uniform mesh $\xi_s = s\Delta, \ s = 0,..,S, \ \Delta = 1/S$

$$u_{s-1}^{\binom{1}{2}} - \left(2 + \overline{\gamma}_{m_0}^2 \Delta^2 \right) u_s^{\binom{1}{2}} + u_{s+1}^{\binom{1}{2}} = -\frac{\Delta^2 2}{\sigma_{n_0}} \cos\{\pi n_0 \Delta s\} \begin{cases} -1 \\ (-1)^{n_0} \end{cases} \quad (1.91)$$

Second-order accurate discretization of boundary conditions (1.76) and (1.77) are

$$u_1^{\binom{1}{2}} = u_0^{\binom{1}{2}} \alpha_0 + \beta_0^{\binom{1}{2}} \quad (1.92)$$

$$\alpha_0 = \left\{ 1 + \frac{1}{2} \overline{\gamma}_{m_0}^2 \Delta^2 \right\}$$

$$\beta_0^{\binom{1}{2}} = \begin{cases} -\Delta + \dfrac{\Delta^2}{\sigma_{n_0}} \\ -\dfrac{(-1)^{n_0} \Delta^2}{\sigma_{n_0}} \end{cases} \quad (1.93)$$

$$u_{S-1} = u_S \alpha_S + \beta_S \quad (1.94)$$

$$\alpha_S = \left\{ 1 + \frac{1}{2} \overline{\gamma}_{m_0}^2 \Delta^2 \right\}$$

$$\beta_S^{\binom{1}{2}} = \begin{cases} \dfrac{(-1)^{n_0} \Delta^2}{\sigma_{n_0}} \\ \Delta - \dfrac{\Delta^2}{\sigma_{n_0}} \end{cases} \quad (1.95)$$

We used a matrix method for solving the two-point boundary value problems (1.91).

Using the above results, we can write the equation (1.78) in the form

$$u_S^{(1)} C^{(k-1,l)} + \left\{ u_S^{(2)} - u_0^{(1)} \right\} C^{(k,l)} - u_0^{(2)} C^{(k+1,l)} = \delta_{k,l} - (-1)^{n_0} \delta_{k-1,l} \quad (1.96)$$

This infinitive system of linear equations is to be truncated under numerical solution of the problem. Taking into account (1.90), we can rewrite this system as

$$\left\{u_S^{(2)} - u_0^{(1)}\right\} C^{(-N+1,0)} - u_0^{(2)} C^{(-N+2,0)} = 0$$

$$u_S^{(1)} C^{(-N+1,0)} + \left\{u_S^{(2)} - u_0^{(1)}\right\} C^{(-N+2,0)} - u_0^{(2)} C^{(-N+3,0)} = 0$$

.....................

$$u_S^{(1)} C^{(-1,0)} + \left\{u_S^{(2)} - u_0^{(1)}\right\} C^{(0,0)} - u_0^{(2)} C^{(1,0)} = 1$$

$$u_S^{(1)} C^{(0,0)} + \left\{u_S^{(2)} - u_0^{(1)}\right\} C^{(1,0)} - u_0^{(2)} C^{(2,0)} = -(-1)^{n_0} \quad (1.97)$$

....................

$$u_S^{(1)} C^{(N-2,0)} + \left\{u_S^{(2)} - u_0^{(1)}\right\} C^{(N-1,0)} - u_0^{(2)} C^{(N,0)} = 0$$

$$u_S^{(1)} C^{(N-1,0)} + \left\{u_S^{(2)} - u_0^{(1)}\right\} C^{(N,0)} = 0$$

The coupling coefficients for the volume with number 0 are

$$\alpha_{m_0,n_0}^{(0,l')} = \frac{2c^2}{\varepsilon \sigma_{n_0} d \, \omega_{m_0,n_0}^2} \left( C^{(-l',0)} - (-1)^{n_0} C^{(-l'+1,0)} \right) \quad (1.98)$$

where $l' = -(N-1),...,0,1,....,(N-1)$

We made several calculations to illustrate the possibilities of the proposed method in the framework of 1-D approach. In Fig. 4-Fig. 8 results are shown.

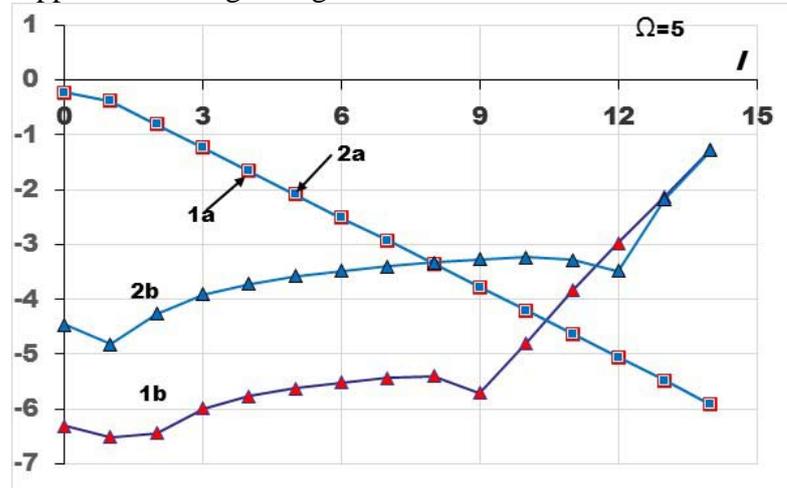

Fig. 4 The coupling coefficients ($\left( \log_{10} \left\{ abs\left( \alpha_{10}^{(0,l)} \right) \right\} \right)$ of the 0-volume with the $l$ volume ($l = 0,1,2...,N-1$) These coefficients were calculated on the base of the two-point boundary value problems (1.91)), 1$a$ - $S$ =1000, 2$a$ - $S$ =100. Errors of calculation of the coupling coefficients ($\log_{10} \left\{ abs\left( \alpha_{10}^{(k,k+l)} / \bar{\alpha}_{10}^{(k,k+l)} \right) \right\}$, $\bar{\alpha}_{10}^{(k,k+l)}$ - the exact solution (1.41)), 1$b$ - $S$ =1000, 2$b$ - $S$ =100. $N$ =15, $\Omega = \dfrac{\omega b}{c} = 5$, $m_0$ =1 and $n_0$ =0, $b/d$ =2, $\varepsilon = 1 - 1E - 6i$

From Fig. 4 it follows that in the case when $|p_2|$ is not close to 1 (see Fig. 2, $\Omega$ =5) the coupling coefficients decrease fast enough with increasing $l$ and the errors of its calculation can be made small enough.
13

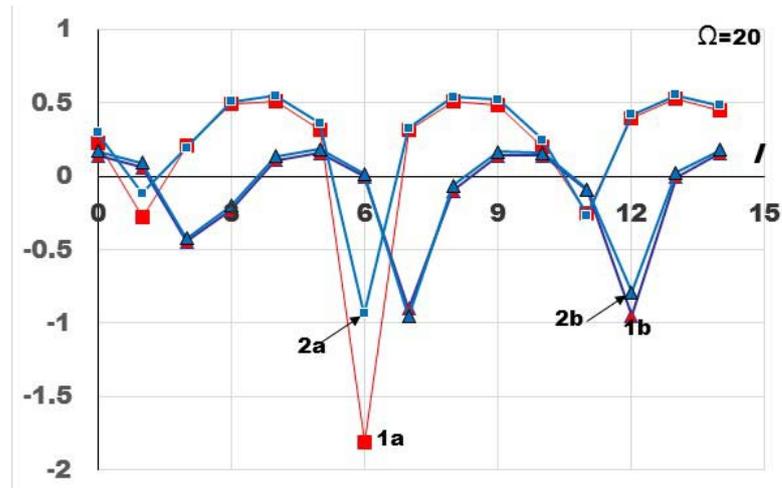

Fig. 5 The coupling coefficients ($\left(\log_{10}\left\{abs\left(\alpha_{10}^{(0,l)}\right)\right\}\right)$ of the 0-volume with the $l$ volume ($l = 0,1,2..., N-1$) These coefficients were calculated on the base of the two-point boundary value problems (1.91)), 1$a$ - $S$ =1000, 2$a$ - $S$ =100. Errors of calculation of the coupling coefficients ($\log_{10}\left\{abs\left(\alpha_{10}^{(k,k+l)} / \overline{\alpha}_{10}^{(k,k+l)}\right)\right\}$, $\overline{\alpha}_{10}^{(k,k+l)}$ - the exact solution (1.41)), 1$b$ - $S$ =1000, 2$b$ - $S$ =100. $N$ =15, $\Omega = \dfrac{\omega b}{c} = 20$, $m_0$ =1 and $n_0$ =0, $b/d$ =2, $\varepsilon = 1 - 1E - 6i$

In the case when $|p_2|$ is close to 1 (see Fig. 2, $\Omega$=20) the coupling coefficients decrease slow[5] with increasing $l$ and the errors of its calculation is not small (see Fig. 5). Increasing the size of matrix in (1.97) do not improve converging (Fig. 6). For the case $\Omega$=5 increasing the size of matrix in (1.97) shows (Fig. 7) that decreasing of the coupling coefficients take place for all $l$ (Fig. 7), but accuracy do not improved (Fig. 8).

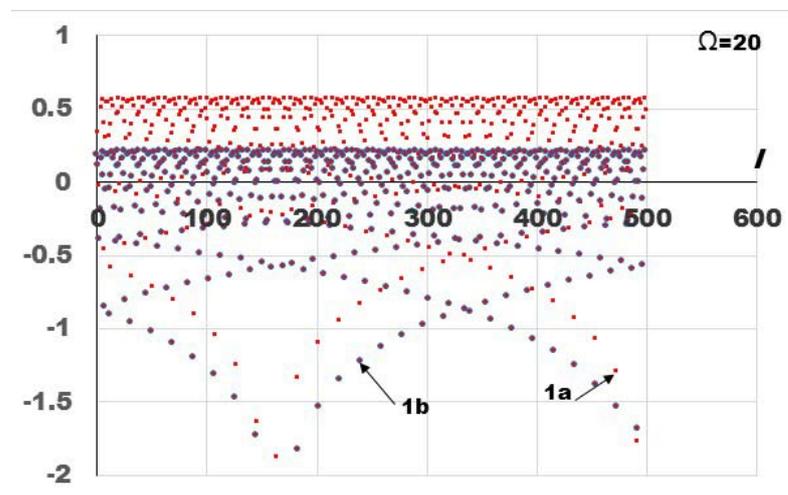

Fig. 6 The coupling coefficients ($\left(\log_{10}\left\{abs\left(\alpha_{10}^{(0,l)}\right)\right\}\right)$ of the 0-volume with the $l$ volume ($l = 0,1,2..., N-1$) These coefficients were calculated on the base of the two-point boundary value problems (1.91)), 1$a$ - $S$ =1000, 2$a$ - $S$ =100. Errors of calculation of the coupling coefficients ($\log_{10}\left\{abs\left(\alpha_{10}^{(k,k+l)} / \overline{\alpha}_{10}^{(k,k+l)}\right)\right\}$, $\overline{\alpha}_{10}^{(k,k+l)}$ - the exact solution (1.41)), 1$b$ - $S$ =1000, 2$b$ - $S$ =100. $N$ =500, $\Omega = \dfrac{\omega b}{c} = 20$, $m_0$ =1 and $n_0$ =1, $b/d$ =2, $\varepsilon = 1 - 1E - 6i$

---

[5] In this case the difference ($|p_2|$ -1) proportional to the imaginary part of $\varepsilon$

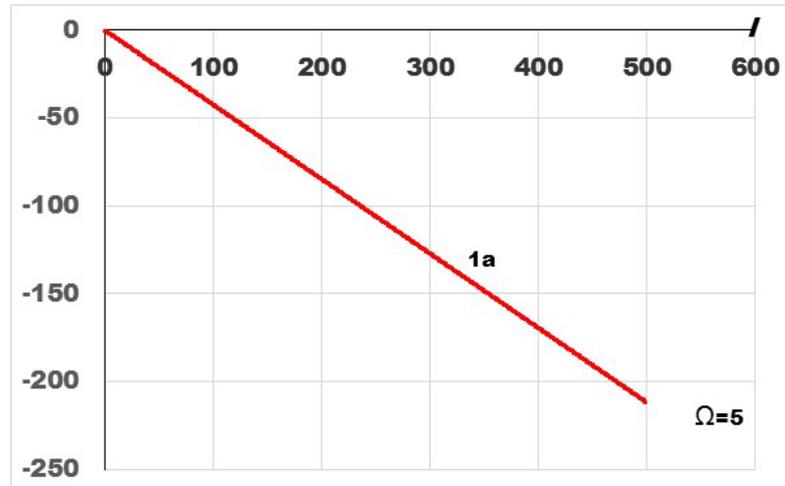

Fig. 7 The coupling coefficients ($\log_{10}\{abs(\alpha_{10}^{(0,l)})\}$) of the 0-volume with the $l$ volume, 1a - $S=1000, N=500$, $\Omega=\dfrac{\omega b}{c}=5$, $m_0=1$ and $n_0=1$, $b/d=2$, $\varepsilon=1-1E-6i$

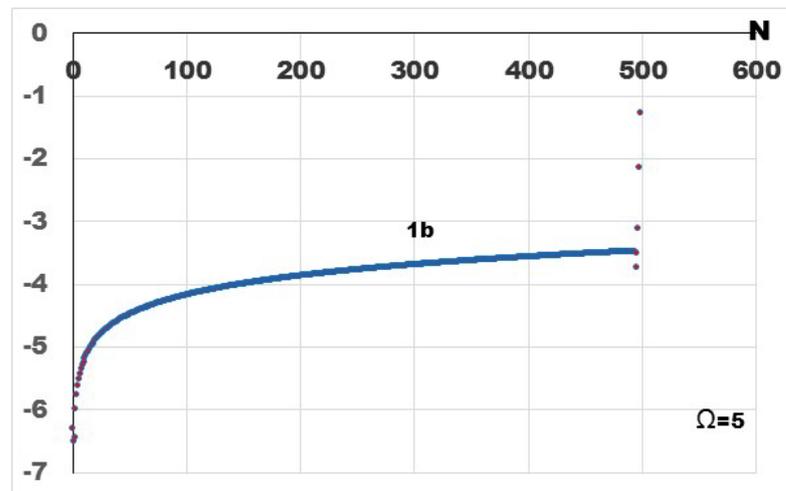

Fig. 8 Errors of calculation of the coupling coefficients ($\log_{10}\{abs(\alpha_{10}^{(k,k+l)}/\bar{\alpha}_{10}^{(k,k+l)})\}$, $\bar{\alpha}_{10}^{(k,k+l)}$ - the exact solution (1.41)), 1b - $S=1000$ $N=500$, $\Omega=\dfrac{\omega b}{c}=5$, $m_0=1$ and $n_0=1$, $b/d=2$, $\varepsilon=1-1E-6i$

### Conclusions

We obtain some results that can be useful in the process of developing the numerical procedure for calculation of the coupling coefficients in the Coupling Cavity Model of arbitrary chain of resonators.